\documentclass[conference]{IEEEtran}
\IEEEoverridecommandlockouts
\usepackage{cite}
\usepackage{amsmath,amssymb,amsfonts}
\usepackage{algorithmic}
\usepackage{graphicx}
\usepackage{textcomp}
\usepackage{xcolor}
\usepackage{url}

\def\BibTeX{{\rm B\kern-.05em{\sc i\kern-.025em b}\kern-.08em
    T\kern-.1667em\lower.7ex\hbox{E}\kern-.125emX}}
\begin{document}

\title{Analysis of User Dwell Time by Category \\ in News Application}

\author{\IEEEauthorblockN{Yoshifumi Seki}
\IEEEauthorblockA{\textit{Gunosy Inc.}\\
Minato-ku, Tokyo, Japan \\
yoshifumi.seki@gunosy.com}
\and
\IEEEauthorblockN{Mitsuo Yoshida}
\IEEEauthorblockA{\textit{Toyohashi University of Technorogy}\\
Toyohashi, Aichi, Japan \\
yoshida@cs.tut.ac.jp}
}

\maketitle

\begin{abstract}
Dwell time indicates how long a user looked at a page,
and this is used especially in fields where ratings from users such as search engines, recommender systems, and advertisements are important.
Despite the importance of this index, however, its characteristics are not well known.
In this paper, we analyze the dwell time of news pages according to category in smartphone application.
Our aim is to clarify the characteristics of dwell time and the relation between length of news page and dwell time, for each category.
The results indicated different dwell time trends for each category.
For example, the social category had fewer news pages with shorter dwell time than peaks, compared to other categories,
and there were a few news pages with remarkably short dwell time.
We also found a large difference by category in the correlation value between dwell time and length of news page.
Specifically, political news had the highest correlation value and technology news had the lowest.
In addition, we found that a user tends to get sufficient information about the news content from the news title in short dwell times.
\end{abstract}


\section{Introduction}

The most widely known index in webpage rating is page view,
because increasing page view generically leads to an increase in revenue from advertisements.
However, there are limits to this index.
Dwell time, on the other hand, is a well-known indicator that is also used as an index for webpage rating~\cite{O'Brien2010}.
Dwell time indicates how long a user looked at a page,
and it is used especially in fields where ratings from users such as search engines~\cite{Agichtein2006,Morita1994}, recommender systems~\cite{Yi2014}, and advertisements~\cite{Lalmas2015,Zhou2016} are important.
Despite the importance of this index, however, its characteristics are not well known.
In addition, previous studies have reported that trends vary greatly between PCs and smartphones~\cite{Lalmas2015,Yi2014},
but there has not been much analysis focusing on smartphones.

In this study, we analyze the characteristics of the dwell time of news pages in smartphone applications according to category.
Due to growing interest in the social issues of fake news~\cite{Allcott2017,Bourgonje2017} and clickbait news~\cite{Chakraborty2016,Potthast2016}, the question of quality of news on the web is becoming more important.
An increasing number of users are concerned about the quality of news pages on the web,
and news media publishers have begun to rely on dwell times to indicate the high value of their media\footnote{Digiday (2017): How The New York Times gets people to spend 5 minutes per visit on its site.}.
However, in order to discuss the quality of news pages based on dwell time,
we need to understand the characteristics of dwell time in news pages, but there is not enough research on this subject.
For example, Yi et al.~\cite{Yi2014} visualize trends in dwell time in the news recommender system,
but they show results for only some categories and do not conduct an overall analysis.
We show the characteristics of the dwell time of news pages using actual data in order to create a basis for discussing the quality of news pages in relation to user behavior such as dwell times.

We focus on the topic category of news when analyzing dwell time.
News categories are important, as Bandari et al.~\cite{Bandari2012} reported that category was the most important feature when predicting the number of news clicks.
In addition, Liu et al.~\cite{Liu2010} report that the variation in parameter for each category is large when modeling dwell time.
Lagun and Lalmas~\cite{Lagun2016} predicted dwell time by extending Latent Dirichlet Allocation (LDA)~\cite{Blei2003}.
Latent topics in LDA are also known to represent categories of documents.
It is suggested that there is a strong relationship between dwell time and category in this way,
but the nature of this relationship has not been sufficiently discussed.

In this paper, we analyze the dwell time of news pages according to category in smartphone application.
Our aim is to clarify the characteristics of dwell time and the relation between length of news page and dwell time, for each category.
To this end, we address the following three research questions:
\begin{description}
 \item[RQ1] Dwell time by category:\\ Is there a difference in the dwell time for each category?
 \item[RQ2] Dwell time and length of news page:\\ Is there a correlation between dwell time and length of news page?
 \item[RQ3] News content with short dwell time:\\ What kind of content on is a news page with a short dwell time?
\end{description}
The results indicated different dwell time trends for each category.
For example, the social category had fewer news pages with shorter dwell time than peaks, compared to other categories,
and there were a few news pages with remarkably short dwell time.
We also found a large difference by category in the correlation value between dwell time and length of news page.
Specifically, political news had the highest correlation value and technology news had the lowest.
In addition, we found that a user tends to get sufficient information about the news content from the news title in short dwell times.

\section{Data}

In this study, we used the browsing data of news pages from July 1 to 7, 2017.
This data was gathered using Gunosy\footnote{\url{http://gunosy.co.jp/en/}}, an information curation service for smartphones.
Gunosy is one of the well-known famous news applications for iOS and Android in Japan.

Our data consists of two types of data: news page data and user behavior data for news pages.
The news page data includes categories, titles, news articles, and thumbnail images.
In Gunosy, news pages are displayed by category which is determined by several heuristic rules and supervised machine learning.
In this study, we used news pages from eight categories: ``politics'', ``economy'', ``society'', ``international'', ``technology'', ``sports'', ``entertainment'', and ``column''.
The user behavior data includes the dwell time on the target news page and the length of the news page.
If a user browses the same news page more than once, the maximum dwell time is taken as the representative value of the user.
The length of the news page is included in the user behavior data because the length varies depending on the device used by the user.

Dwell time should be related to news page.
We regard the dwell time of a news page as the median of the dwell time of the user who viewed the news page.
If the user leaves the application launched, a very long dwell time will be recorded.
Therefore, it is not appropriate to use the average value as a representative value of dwell time.

We analyze the dwell time of news pages, but this may be an unreliable measure for pages that are rarely viewed.
Also, the page view tendency differs for each category, and there are categories that are frequently viewed and those that are rarely viewed.
Therefore, only news pages in the top 10\% of page view for each category are used.

\section{Analysis and Results}

\subsection{Dwell time by category}

The length of dwell time has been used as an index for rating web pages.
Previous studies have suggested a strong relationship between dwell time and category,
but the nature of this relationship has not been sufficiently discussed.
This section addresses the following research question:
Is there a difference in the dwell time for each category?

\begin{figure}[tp]
  \centering
  \includegraphics[width=0.7\linewidth]{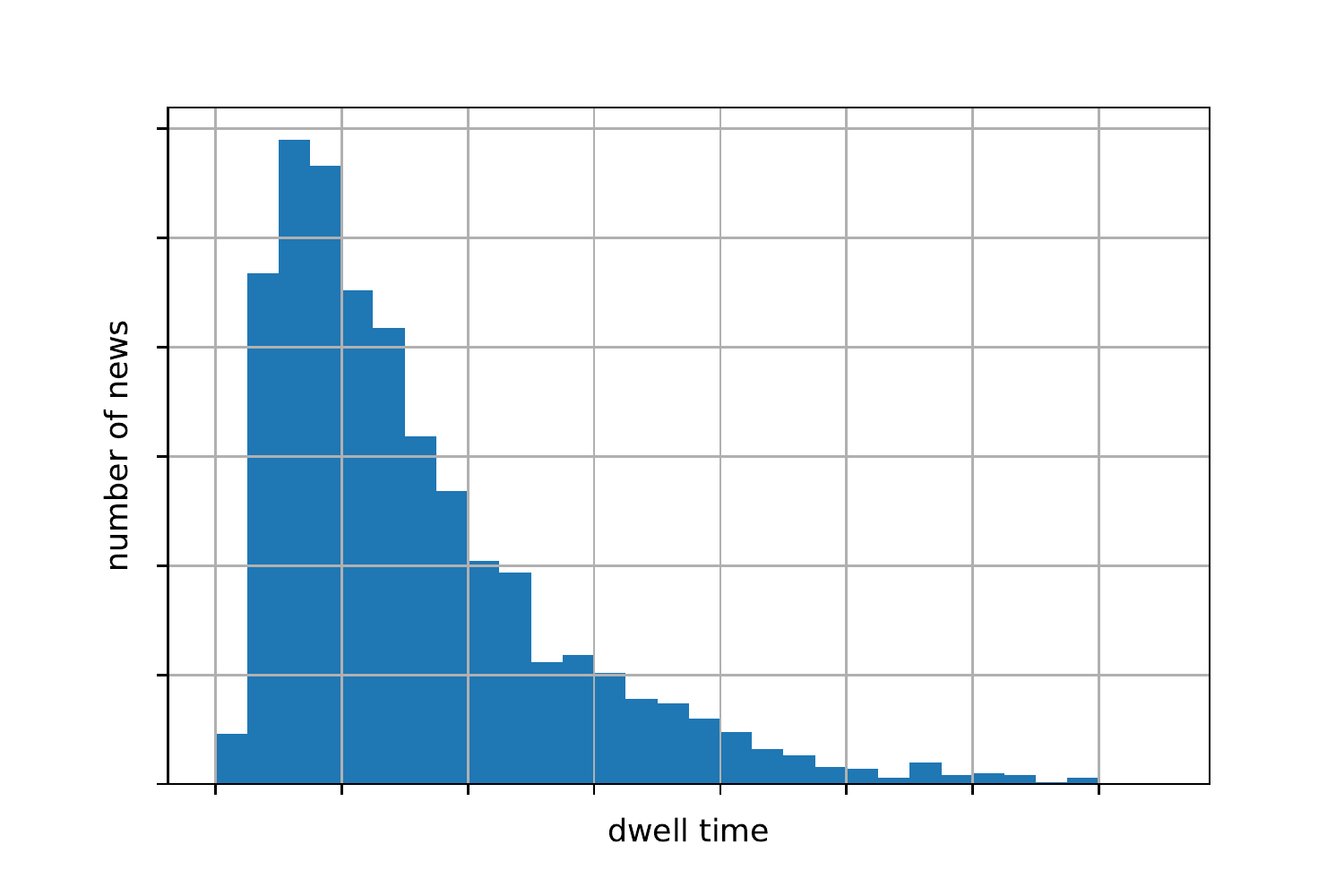}
  \caption{Histogram of dwell time: The x-axis indicates the dwell time by news page. The y-axis indicates the number of such news pages. There are many news pages with a relatively short dwell time.}
  \label{fig:dwell_time}
\end{figure}

A histogram of the dwell time of all news pages is shown in Fig.~\ref{fig:dwell_time}.
The values of the x-axis and the y-axis are equally spaced~\footnote{Due to limitations because of the Gunosy privacy policy, we discussed using relative values instead of absolute values.}.
The visualization does not include e do not use news pages whose dwell time is over a certain threshold for visualization,
but these excluded news pages are account for less than 0.5\% of the total.
There is a peak at a relatively short dwell time, and the frequency decreases from there.

\begin{figure}[tp]
  \includegraphics[width=0.99\linewidth]{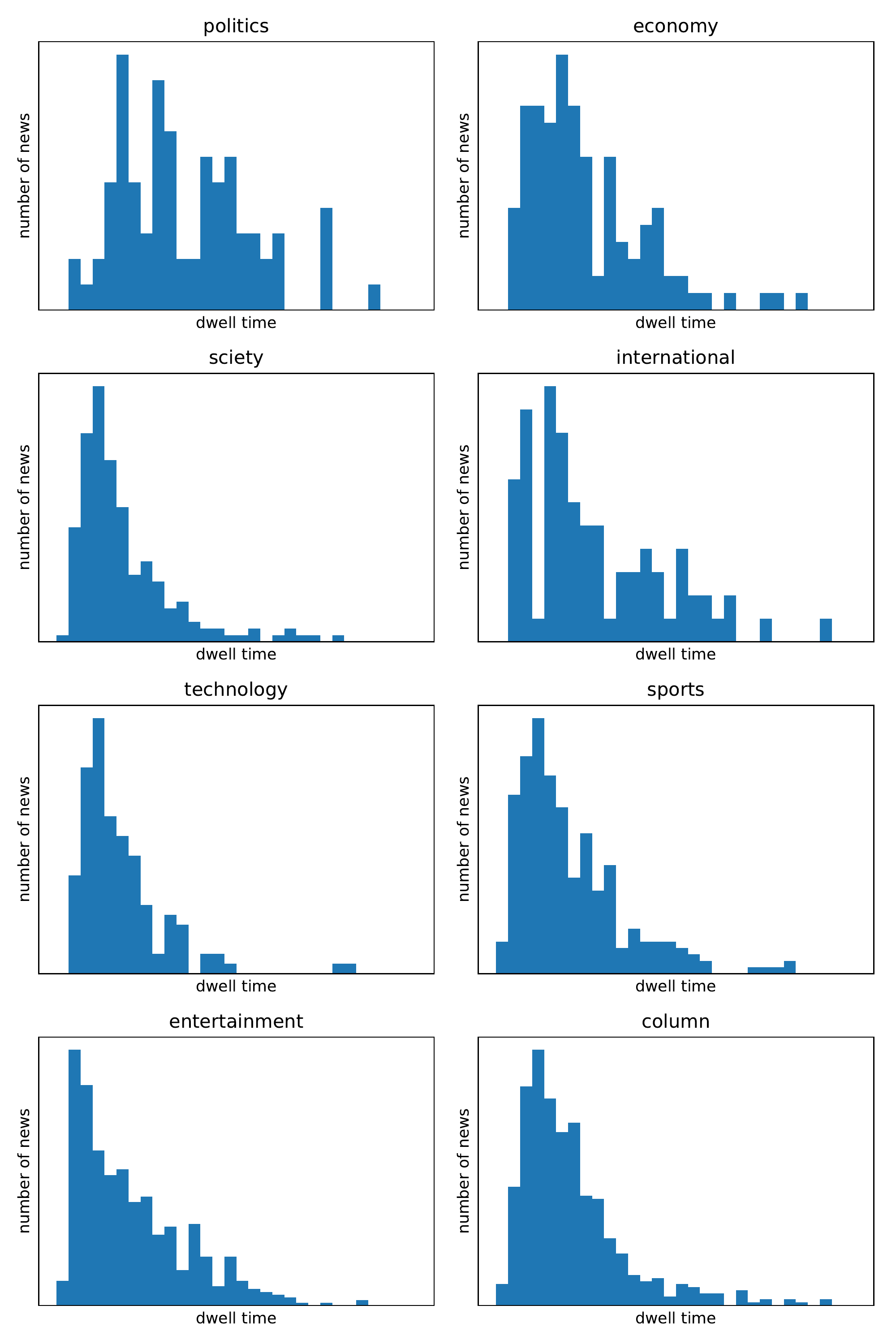}
  \caption{Histogram of dwell time by category: The x-axis indicates the dwell time by news page. The y-axis indicates the number of such news pages. Each category has a different tendency.}
  \label{fig:all_category_dwell_time}
\end{figure}

Fig.~\ref{fig:all_category_dwell_time} shows the histogram of dwell time by category.
The x-axis is common to all categories, as shown in Fig.~\ref{fig:dwell_time}.
The y-axis differs according to the maximum frequency of each category.
The peak of thedwell time in the economy category in Fig.~\ref{fig:all_category_dwell_time} is short compared to the histogram in Fig.~\ref{fig:dwell_time},
but the decay of the value is not steep, and there are a lot of news pages with relatively long dwell times.
The social category has fewer news pages with shorter dwell time than peaks, compared to other categories,
and there are a few news pages with remarkably short dwell time.
The entertainment category tends to have many more news pages with shorter dwell times than other categories do.
The column category is similar in tendency to the social category, but the decay of the value is not steep.
As described above, the characteristics of the histogram are different for each category.

\subsection{Dwell time and length of news page}

Assuming that the user reads all news pages to the end, the dwell time correlates with the length of the news page.
This section addresses the following research question:
Is there a correlation between dwell time and length of news page?
To confirm this, we obtain the correlation between the dwell time and the length of news page for each category by using Pearson's correlation coefficient.

The results are shown in TABLE~\ref{tbl:coherence_dwell_and_length}.
The correlation coefficient of the whole is not as high as about 0.3.
This is dragged by correlation coefficients in some categories because the number of news pages has a bias for each category.
By category, the correlation coefficients of politics, international, and sports are as high as about 0.55,
and there is a relatively strong correlation between dwell time and length of news page.
On the other hand, the correlation coefficients of technology, entertainment and column are as low as about 0.3,
and the correlation between dwell time and length of news page is low.
In this way, correlation coefficients have a bias for each category.

\begin{table}[t]
  \caption{Correlation coefficient between dwell time and the length of news page: The values for politics and sports are as high as 0.55 or more, but the value for technology is as low as below 0.25.}
  \centering
  \begin{tabular}{l|rr}
    \hline
    \hline
    \textbf{Category} & \textbf{Correlation coefficient} & \textbf{\# of news} \\
    \hline
politics & 0.602 & 4.05\% \\
economy & 0.432 & 5.72\% \\
society & 0.454 & 9.52\% \\
international & 0.527 & 4.05\% \\
technology & 0.245 & 6.22\% \\
sports & 0.561 & 13.42\% \\
entertainment & 0.309 & 29.26\% \\
column & 0.367 & 27.78\% \\
    \hline
    \textbf{whole} & 0.291 & \\
    \hline
  \end{tabular}
  \label{tbl:coherence_dwell_and_length}
\end{table}

\subsection{News content with short dwell time}

The previous section showed that there was not much correlation between dwell time and length of news page.
We also found a different trend for each category.
In this section, we focus on dwell time and address the following research question:
what kind of content on is a news page with a short dwell time?

On news pages with short dwell time, was the user not satisfied?
In previous studies, the answer to this question has been ``No, I am not satisfied.''
However, a user can understand the content from the title of the news page and may not have read the news text.
In addition, we focus on the length of the news page, not the length of the news text.
Photos increase the length of the news page, but it takes no time to read a photo.
We hypothesized that the user is satisfied with the news page even when the dwell time is short.

\begin{figure}[tp]
  \centering
  \includegraphics[width=0.99\linewidth]{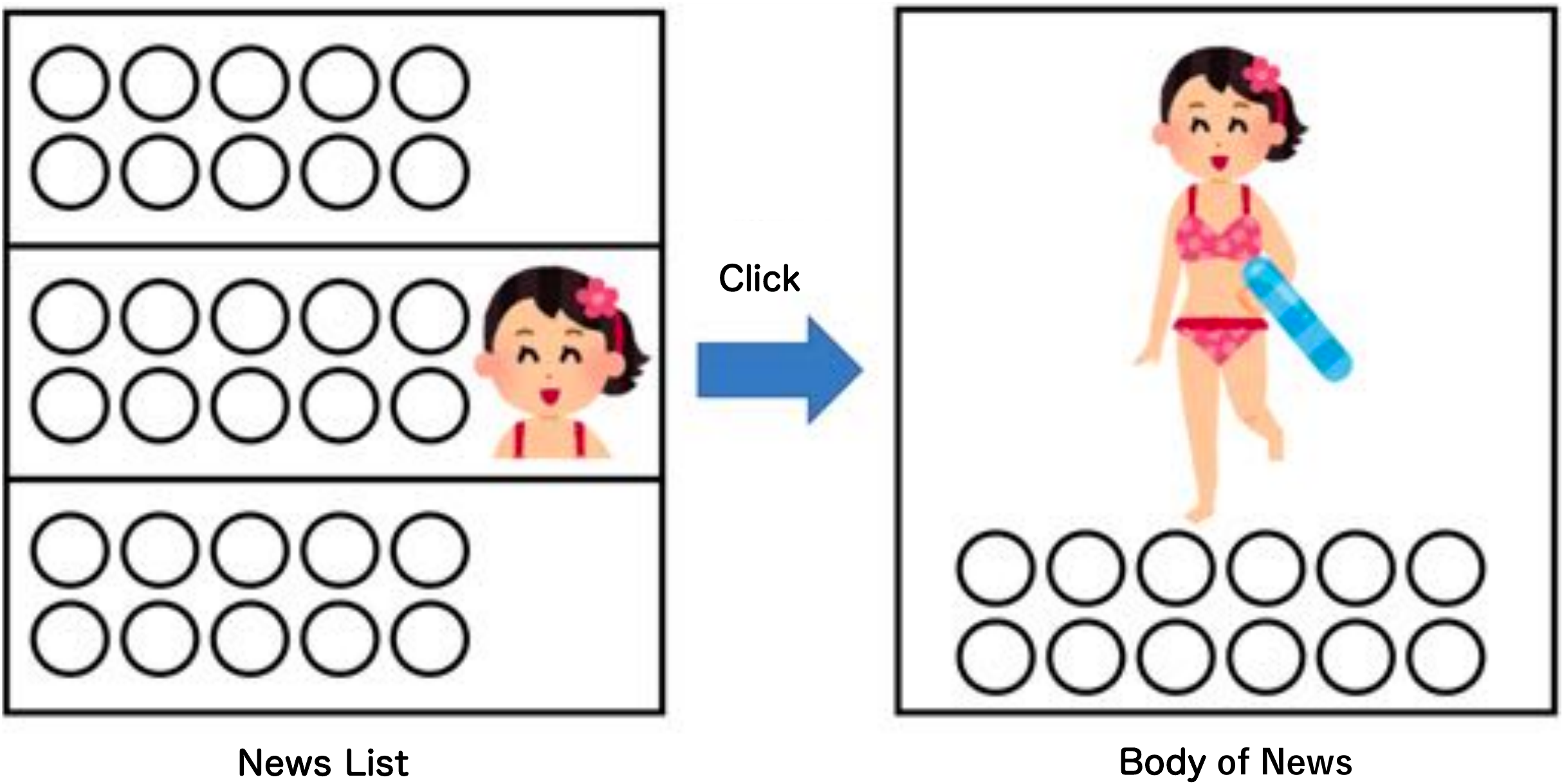}
  \caption{Image with thumbnail photo trimmed: Users who only want to enlarge the photo will have a shorter stay.}
  \label{fig:image_trimming}
\end{figure}

We assume the following about news pages with remarkably short dwell time:
\begin{enumerate}
\renewcommand{\labelenumi}{(\alph{enumi})}
  \item The title has a sufficient amount of information, and the title matches the content of the text.
  \item The information expected from the title is not in the content of the text.
  \item The title recalls photos, and the desire for information is satisfied by viewing large photos.
  \item The title recalls photos, but the expected photos do not appear in the body.
\end{enumerate}
The case of (a) is evident in news stories that simply convey facts, for example, a news article that states, ``Mr. A won in City B mayoral election.''
Since there is not much information in the text, the dwell time is short, but the user will be satisfied with the news page.
In the case of (b), there are many news pages in which only a part of the events are indicated in the title.
For example, in a news article titled ``Mr. A loved Ms. B,''
if the content of this news is a press release of a movie, Mr. A did not actually love Ms. B, but it is a story in a movie.
Such titles are dishonest, and users may be dissatisfied with the news.
The case of (c) is often seen in news pages on album releases with the musician's photo.
In Gunosy, especially, when displaying the thumbnail photo on the news list, as shown in Fig.~\ref{fig:image_trimming},
if the photo is of a person, the face of the person is extracted as the center.
This may motivate a user to view the entire photo.
Since a user only enlarges the image, the dwell time is short, but the user will be satisfied with the news page.
The case of (d) is often seen in news pages on album releases with the musician's photo, as in (c).
A photos of the musician is used in the title.
A user expects such a photo in the body,
but a photo of a press conference has been used in the body instead, and users may be dissatisfied with the news.

We analyzed the news articles in relation to these four cases.
We divided the four cases into groups: (a) and (b) for text and (c) and (d) for photos.
For (a) and (b), it is important whether the title matches the content of the text.
However, there are cases where there is not enough information in the title.
Therefore, we consider whether the title has enough information and whether the title matches the content of the text.
For (c) and (d), we consider whether the title recalls the photo and whether the photo matches the recalled content.

Based on the above, we prepared the following four questions for news pages.
\begin{enumerate}
  \item Does the title have enough information?
  \item Are the title and body sufficiently matched?
  \item Does the title recall photos?
  \item Does a proper photo appear in the body?
\end{enumerate}
The author answered the four questions for news pages whose dwell time is not over a certain limit.
There were 321 target news pages.

The results of this investigation are shown in TABLE~\ref{tbl:all_answers}.
The answer of ``Yes'' to each question holds for news pages that satisfy the user.
About two-thirds of all news pages answer ``Yes,'' and even if the dwell time is short, users may be satisfied with the news page.
In addition, 85\% of news pages recalled photos.
Many news pages with short dwell times had titles to recall photos.

\begin{table}[t]
  \caption{Results of answers to news pages.}
  \centering
  \begin{tabular}{l|r|r}
  \hline
  \hline
  Question & Yes & No \\
  \hline
  1) Does the title have enough information? & 253& 68 \\
  2) Are the title and body sufficiently matched? & 246 & 75 \\
  3) Does the title recall photos? & 273 & 48 \\
  4) Does a proper photo appear in the body? & 209 & 112 \\
  \hline
  \end{tabular}
  \label{tbl:all_answers}
\end{table}

\section{Conclusion}

We analyzed the dwell time of news pages according to category in smartphone application.
Our aim was to clarify the characteristics of dwell time and the relation between length of news page and dwell time, for each category.
To this end, we addressed three research questions.
The results indicated different dwell time trends for each category.
For example, the social category had fewer news pages with shorter dwell time than peaks, compared to other categories,
and there were a few news pages with remarkably short dwell time.
We also found a large difference by category in the correlation value between dwell time and length of news page.
Specifically, political news had the highest correlation value and technology news had the lowest.
In addition, we found that a user tends to get sufficient information about the news content from the news title in short dwell times.

\bibliographystyle{IEEEtran}
\bibliography{IEEEabrv,references}

\end{document}